\documentclass{aa}  

\usepackage{graphicx}
\usepackage{txfonts}
\usepackage{multirow}

\begin{document}

   \title{A long optical plateau in the afterglow of the Extended Emission short GRB 150424A}

   \subtitle{Evidence for energy injection by a magnetar?}

   \author{F. Knust \inst{1}
          \and J. Greiner\inst{1}
          \and H. J. van Eerten \inst{2}
          \and P. Schady \inst{1}
          \and D. A. Kann \inst{3}   
          \and T.-W. Chen \inst{1}
          \and C. Delvaux \inst{1}
          \and J. F. Graham \inst{1}
          \and S. Klose \inst{4}          
          \and T. Krühler \inst{1} 
          \and N. J. McConnell \inst{5}          
          \and A. Nicuesa Guelbenzu \inst{4}
          \and D. A. Perley \inst{6}          
          \and S. Schmidl \inst{4}
          \and T. Schweyer \inst{1}               
          \and M. Tanga \inst{1}
          \and K. Varela \inst{1}
          }

   \institute{Max-Planck-Institut für Extraterrestrische Physik,
                Giessenbachstraße, 85748, Garching, Germany\\ 
               \email{fknust@mpe.mpg.de}
         \and 
              University of Bath,
              Bath BA2 7AY, United Kingdom
         \and 
              Instituto de Astrof\'isica de Andaluc\'ia (IAA-CSIC),
              Glorieta de la Astronom\'ia s/n, 18008 Granada, Spain
         \and
              Thüringer Landessternwarte Tautenburg,
               Sternwarte 5, 07778 Tautenburg, Germany
         \and
             University of California, 1156 High Street, Santa Cruz, Ca 95064, US
         \and
             Astrophysics Research Institute, 
             Liverpool John Moores University, 
             IC2, Liverpool Science Park, 146 Brownlow Hill, Liverpool L3 5RF, UK
             }


 
  \abstract
   {Short-duration GRBs with extended emission form a subclass of short GRBs,
   comprising about 15\% of the short-duration sample.
   Afterglow detections of short GRBs are also rare (about 30\%) due to 
   their smaller luminosity.
  }   
   {We present a multi-band data set of the short burst with extended emission
    GRB 150424A, comprising of  GROND observations, complemented with data 
   from \emph{Swift}/UVOT, \emph{Swift}/XRT,  {HST, Keck/LRIS} and  data points from
   the literature. The GRB 150424A afterglow shows an extended plateau phase,
   lasting about 8hrs. The analysis of this unique GRB afterglow might shed
   light on the understanding of afterglow plateau emission, the nature of
   which is still under debate. 
   }
   {We present a phenomenological analysis by applying fireball closure 
    relations, and interpret the findings in the context of the fireball model. 
    We discuss the plausibility of a magnetar as a central engine, being 
    responsible for additional and prolonged energy injection into the fireball.
   }   
   {We find convincing evidence for energy injection into the afterglow of
   GRB 150424A.
   We find that a magnetar spin down as source for a prolonged energy injection
   requires that at least 4\% of the spin-down energy is converted to radiation.
   }
   {}


   \maketitle
%

\section{Introduction}
Gamma Ray Bursts (GRBs) are among the most luminous explosions in the Universe.
They are characterized by an intense prompt $\gamma$-ray flash, followed by a broad-band afterglow. 
We distinguish between two flavours: \textit{short duration GRBs} and \textit{long duration GRBs} \citep{Kouveliotou1993}.
Some of those short GRBs show Extended Emission (EE) in the $\gamma$ band, after the short $\gamma$ flash (e.g. \citealt{Mazets2002, Norris2006, Norris2010}).
While classically GRBs that emit 90\% of their prompt emission energy in $T_{90} \lesssim 2$s are classified as short, 
EE short GRBs can have a significantly longer $T_{90}$.
Since EE is also spectrally softer, the classification according to $T_{90}$ is debated. 

 Short GRBs are generally about a factor of 10-100 less energetic than the more common long GRBs \citep{Ghirlanda2009} 
and also have a fainter afterglows  \citep{Berger2007, Berger2010, Nakar2007, Gehrels2008, Nysewander2009, Kann2011}, which makes follow up observations challenging.

Short GRBs are believed to originate from older stellar populations (for reviews, see e.g. \citealt{Fong2013, Berger2014}), 
and often occur at a relative offset to the host galaxy's center \citep{Belczynski2006, Fong2010, Church2011, Fong2013a, Behroozi2014}.
Afterglow analysis in the context of the "fireball model" (\citealt{Meszaros1997}; for reviews, see in e.g. \citealt{Piran2004, vanEerten2015}) implies that they have a relatively low circumburst density
(median density  $n \approx (3-15) \times 10^{-3}\, \mathrm{cm}^{-3}$, 80\%-95\% of bursts have densities of $n \lesssim 1 cm^{-3}$, \citealt{Fong2015}).

The possible progenitors of short GRBs are still under some debate.
The most favored progenitors today are  compact binary mergers (CBMs). 
CBMs as progenitors are supported by evidence for a binary neutron star merger (kilonova) association \citep{Tanvir2013, Berger2014}.
The short GRB rate also is consistent with the expected CBM rate \citep{Fong2012, Wanderman2015}.
The offsets of short GRBs to the center of their host galaxies correspond  with the theoretical predictions for the kick a compact binary receives when formed \citep{Berger2010}.
Moreover, unlike for long GRBs, core collapse supernova are ruled out due to a lack of observational associations \citep{Hjorth2005a}. 

During the merging process, the two neutron stars can either collapse directly into a black hole,
 or form a strongly magnetized and rapidly rotating neutron star: a magnetar \citep{Duncan1996, Yi1997, Metzger2008, Zhang2001}.
A magnetar would lose energy via dipole radiation and could provide a prolonged energy injection into the GRB blast wave.
This energy injection would explain a "plateau phase".
A plateau phase is a shallow decay phase in the afterglow light curve with a temporal slope $\alpha \lesssim 1/4$
\footnote{The sign of the slope is convention. We follow Eq. \ref{eq:ClosureRelations}, where a positive number means a decaying light curve.}
, where the standard model predicts a temporal slope $\alpha \sim 1$.
However, the local  magnetar rate is not in agreement with the GRB rate  \citep{Rea2015}. 

Since the first optical detection of an afterglow from a short GRB \citep{Hjorth2005}, fewer than 90 short GRB X-ray afterglows have been detected, as compared to around 1000 long GRB afterglows
\footnote{http://www.mpe.mpg.de/~jcg/grbgen.html}.
Just about 30\% of all short GRB afterglows had an optical/near infrared counterpart \citep{Fong2015}.
Among those afterglows, just a few show a plateau phase:
GRB 060313A \citep{Roming2006};
GRB 061201A \citep{Stratta2007};
GRB 130603B \citep{Fan2013, deUgartePostigo2014a};
see also \citet{Kann2011} for GRB 060313A, GRB 061201A, GRB 090510A).

In this work we analyze afterglow data of GRB 150424A: An EE short GRB with early multi-band coverage.
We use data from the Gamma-ray burst Optical Near-infrared Detector (GROND) \citep{Greiner2008}, 
 the Swift/Ultraviolet and Optical Telescope (UVOT) \citep{Roming2005} and the Swift/X-Ray Telescope (XRT) \citep{Burrows2005}, 
  {the Hubble Space Telescope (HST) and Keck/LRIS}.
 {This high quality data-set makes GRB 150424A one of the best-detected EE short GRBs with an optical plateau phase.}

In Sec. \ref{sec:Data} we present the data we used.
In Sec. \ref{sec:Analysis} we perform a phenomenological analysis and present its physical implications.
In Sec. \ref{sec:Discussion} we discuss the results and its implications for the physical nature of the GRB, 
followed by our conclusions in Sec. \ref{sec:Conclusions}.

\section{Data}\label{sec:Data}
\begin{figure}[ht]
  \centering
  \includegraphics[width=\hsize]{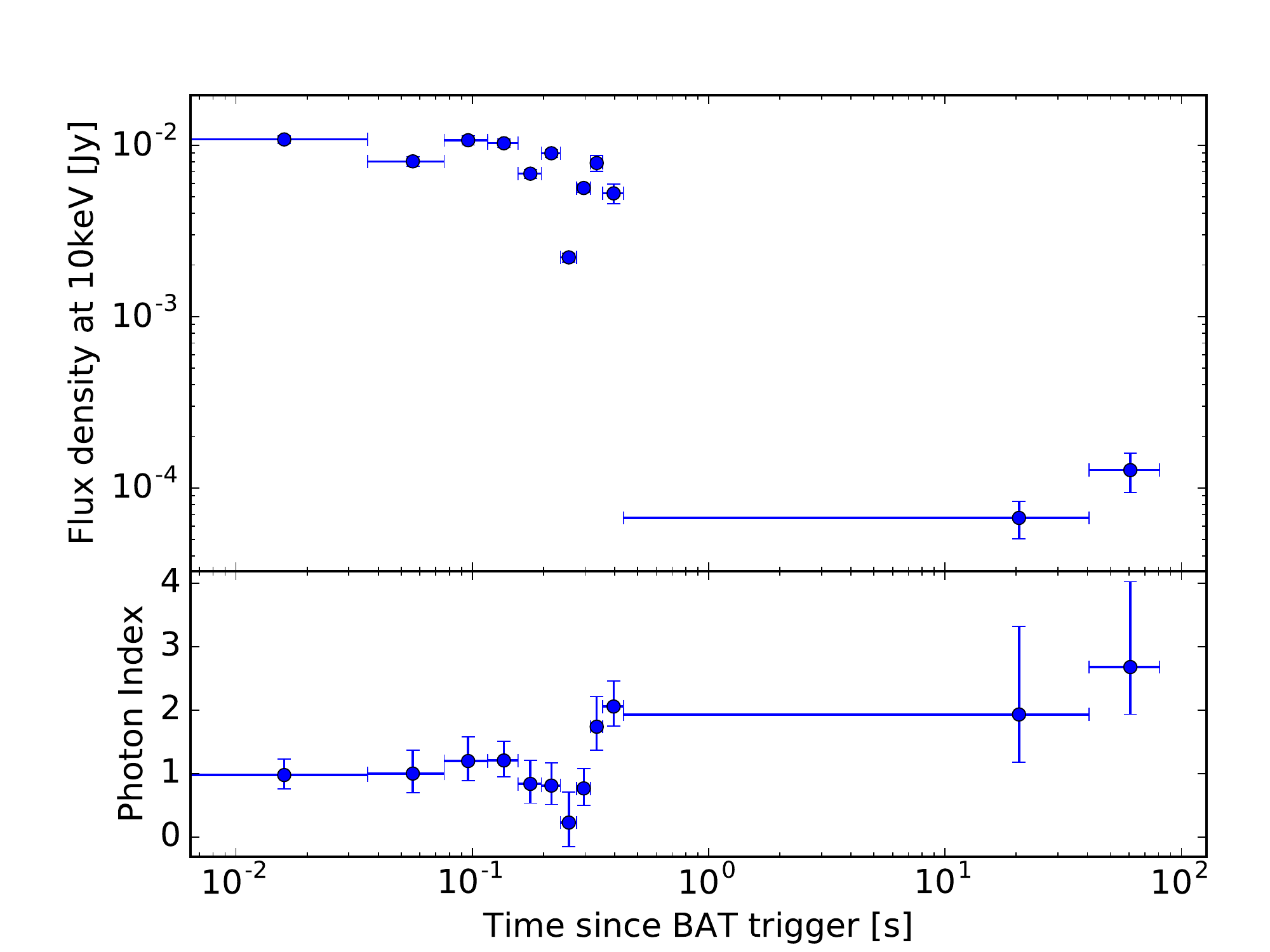}
   \caption{ {Gamma-ray light curve of GRB 150424A. 
   After the initial spike, the GRB shows extended emission for  $\sim 100\,$s after the BAT trigger.} 
   }
   \label{Fig:Gamma_LC}
\end{figure}

On 24th April 2015  {at 07:42:57 UT} the \emph{Swift} Burst Alert Telescope (BAT) detected the short GRB150424A  with a single peak of $0.5\,$s duration.
 {Swift slewed immediately to the burst and a} fading X-ray counterpart was detected by the \emph{Swift}/XRT \citep{Beardmore2015}.
A non fading optical counterpart was found \citep{Marshall2015} with the \emph{Swift}/UVOT \citep{Roming2005}. 
In a refined analysis weak extended emission for $\sim 100\,$s was found \citep{Barthelmy2015} and resulted in a $T_{90}= 91 \pm 22\,$s 
( {the $\gamma$-ray light curve can be seen in Fig. \ref{Fig:Gamma_LC}}).

The UVOT observations cover the time from $82\,\mathrm{s}-1.4 \times 10 ^{6}\,\mathrm{s}$ after the burst.
UVOT photometry was carried out on pipeline processed sky images downloaded from the {\em Swift} data center \footnote{www.swift.ac.uk/swift\_portal} following the standard UVOT procedure \citep{Poole2008}. 
Source photometric measurements were extracted from the UVOT early-time event data and later imaging data files
 using the tool {\sc uvotmaghist} (v1.1) with a circular source extraction region of $5\arcsec$ radius for 
 the first 16ks of data, after which a $3.5\arcsec$ source region radius was used to maximize the signal-to-noise. 
In order to remain compatible with the effective area calibrations, which are based on $5\arcsec$ aperture
photometry, an aperture correction was applied on the photometry that was extracted using the smaller source aperture. 
We consider a signal to noise ratio of $\sigma > 3$ as detection.

1.6 hrs after the burst \citet{Perley2015} reported a g- and R- band detection of the afterglow with the Low Resolution Imaging Spectrometer (LRIS).
GROND was able to follow up \citep{Kann2015} 15 hrs after the burst.
  {GROND detected the afterglow in the $g'$, $r'$, $i'$, $z'$ and $J$ band 
 at RA, Dec.(J2000)= 10:09:13.39, -26:37:51.46 (152.3058, -26.63096) with an uncertainty of $0.3\arcsec$, } 
 and observed strong fading at this point in time. 
All in all, GROND observations covered 8 epochs with detections in the first 3 epochs.
We performed aperture photometry on the LRIS and GROND data, using our IRAF/PyRAF \citep{Tody1993} based pipeline \citep{Yoldacs2008, Kruhler2008},
 and calibrated the data against secondary field stars.

\begin{figure}[ht]
  \centering
  \includegraphics[width=\hsize]{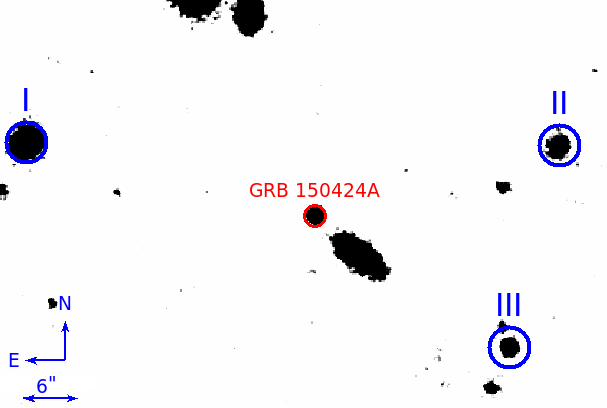}
   \caption{GROND finding chart of GRB 150424A (GROND r band). 
   There is a galaxy south west of the afterglow, but late time HST observations find a weak expanded source at the position of the afterglow
   which is believed to be the host. 
   }
   \label{Fig:150424A_fc}
\end{figure}

The GRB afterglow is north east of an elliptical galaxy with a spectroscopic redshift $z = 0.3$ \citep{Castro-Tirado2015} (see finding chart in Fig. \ref{Fig:150424A_fc}), 
but there are  Hubble Space Telescope (HST) detections \citep{Tanvir2015}  of an extended source that is more likely to be the host galaxy at the position of the afterglow.
 {More HST data of the WFC3 instrument was obtained from the data archive at the Space Telescope Science Institute 
\footnote{http://archive.stsci.edu/hst/index.html}.
 We performed aperture photometry using a $0.4\arcsec$ aperture.}

\citet{Fong2015a} report a VLA 9.8GHz detection 18hrs after the trigger, 
and \citet{Kaplan2015} report early  MWA upper limits in the MHz regime.

\section{Analysis}\label{sec:Analysis}
\subsection{Phenomenology}
A GRB afterglow can be described by the empirical flux description
\begin{equation}\label{eq:ClosureRelations}
F_\nu(t) \propto t^{-\alpha} \nu^{-\beta}
\end{equation} 
with time $t$ and observed frequency $\nu$.
The temporal slope $\alpha$ and the spectral slope $\beta$ depend on the observed spectral regime and can change over time
\citep{Meszaros1997, Granot2002}.

\begin{figure*}[ht]
  \centering
  \includegraphics[width=1.1\hsize]{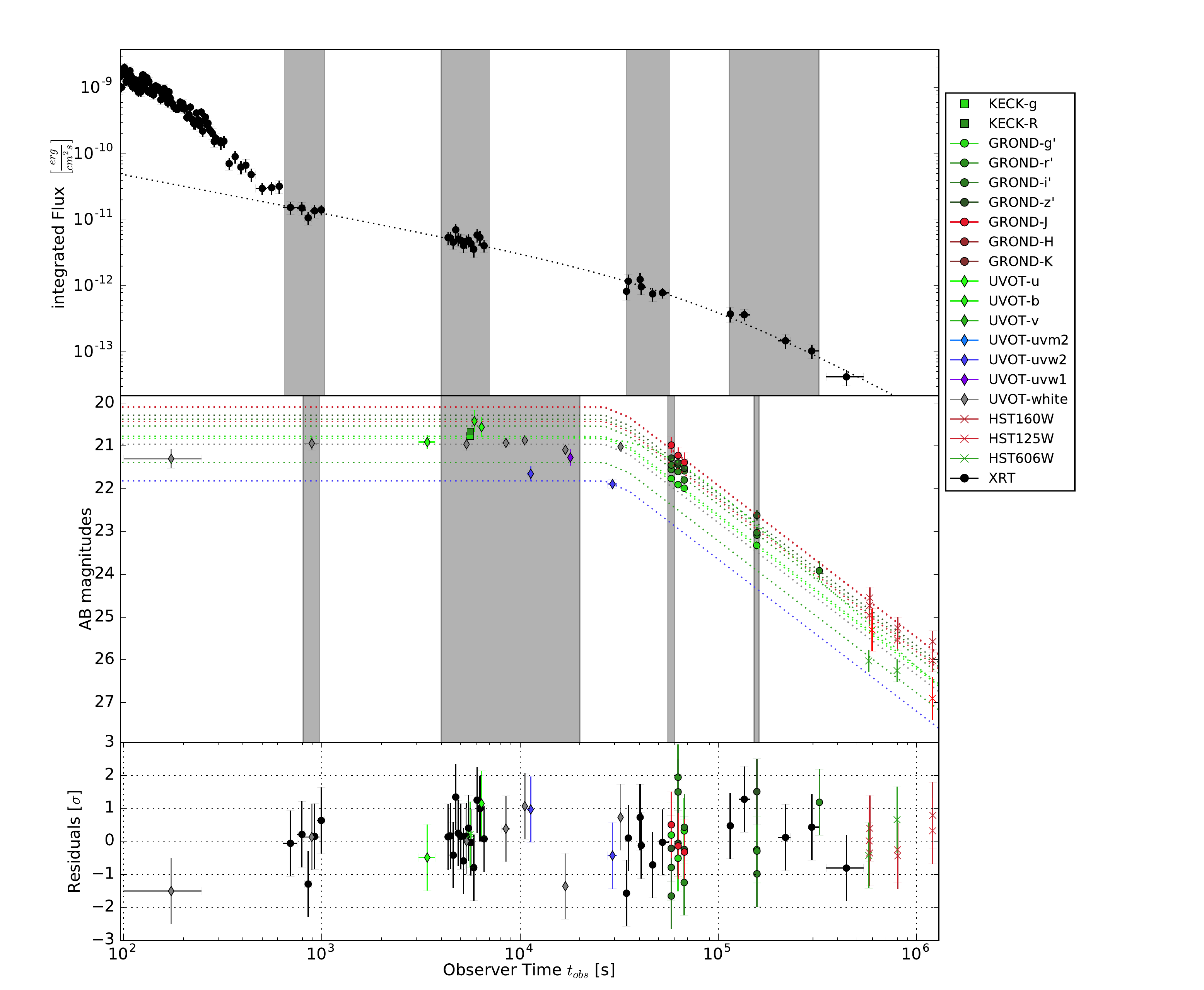}
   \caption{XRT and GROND light curves of GRB 150424A. The shaded areas correspond to the time slices of the snap shot analysis. 
   The dotted lines are the best fitting smoothly broken power laws. 
   The X-ray light curve fit has a reduced $\chi^2= 0.57$, the optical light curve fit has a reduced $\chi^2=1.04$.
   We fit all bands with more than one detections: g', r', i', z', J, white, u, uvw2, HST160W, HST125W and HST606W.
   All detections and upper limits (not shown in light-curve) are given in Tab. \ref{Tab:photometry}.
   The temporal slopes $\alpha$ are given in Tab. \ref{Tab:PhenoAnalysis}.
    }
   \label{Fig:LC150424A}
\end{figure*}

The optical light curve of  GRB 150424A (see Fig. \ref{Fig:LC150424A}) consists of two segments, a plateau and a decay phase.
Both phases are covered by the UVOT observations.
The GROND observations cover the decay phase and constrain its temporal slope. 
We fit all optical bands with more than one detections with one smoothly broken power law per band.
They all share their temporal slopes $\alpha$ and break time $t_{\mathrm{break, opt}}$.

The X-ray light curve shows a steep decay until around $10^{3}$s, followed by a shallow decay phase, steepening again after $\sim 10^{5}\,$s.
The steep decay phase is most likely due to the tail of prompt emission, so 
we ignore the X-ray data before  $10^{3}$s, since a standard fireball afterglow model is not applicable.
The rest of the X-ray light curve we fit with a smoothly broken power law.

To determine the spectral slopes $\beta$,
we perform a joined broad-band fit of optical and X-ray data using {\sc xspec} \citep{Arnaud1996}.
The fireball model assumes synchrotron as the underlying emission mechanism,
therefore the spectral slopes just depend on the electron energy distribution.
The model uses a power law like electron energy distribution with slope $p$.
Assuming a constant $p$, the spectral slopes do not change. 
The light curve fit supports the assumption, 
since thanks to the multi-band capabilities of GROND a chromatic break would be clearly seen in the residuals around the optical break time. 

We choose 4 time slices over which we integrate the XRT spectral counts (indicated with the grey bars in Fig. \ref{Fig:LC150424A}). 
Then we re-bin each spectrum, where we have to find an optimum between counts per spectral bin and number of spectral bins. 
For SED2 and SED3 we re-normalize the total flux of the spectrum to the flux of the XRT light curve at the point in time of the optical data.
For SED0 and SED1 we do not expect the total flux to be significantly different from the flux at the point in time of the optical data.
Afterwards we add a systematic error of 10 \%, accounting for the flux calibration relative to the optical data.

First, we correct the optical/NIR and UV data for galactic extinction $E(B-V)=0.0513 \pm  0.0024$ \citep{Schlafly2011}, then
we use a model consisting of a (broken) power law, 
and involve galactic absorption $N_H = 0.602 \times 10^{21} \mathrm{cm}^{-2}$  \citep{Kalberla2005}, and host extinction and absorption.
We fit the 4 SEDs simultaneously, while each single SED has an individual break frequency and normalization.
We find the spectral slopes before and after the temporal break to be consistent,
so for the final fit we link them for all 4 SEDs.
We also use the constraint $\beta_{\mathrm{optical}}-\beta_{\mathrm{X-ray}}=0.5$ for a synchrotron SED in the slow cooling case with the cooling frequency $\nu_c$ between optical and X-ray.
Not setting this constraint leads to a degeneracy between the spectral slopes and the break frequency. 
The resulting optical slope also is consistent with the fast cooling case where $\beta_{\mathrm{opt}} = 0.5$.

\begin{figure}[ht]
  \centering
  \includegraphics[width=\hsize]{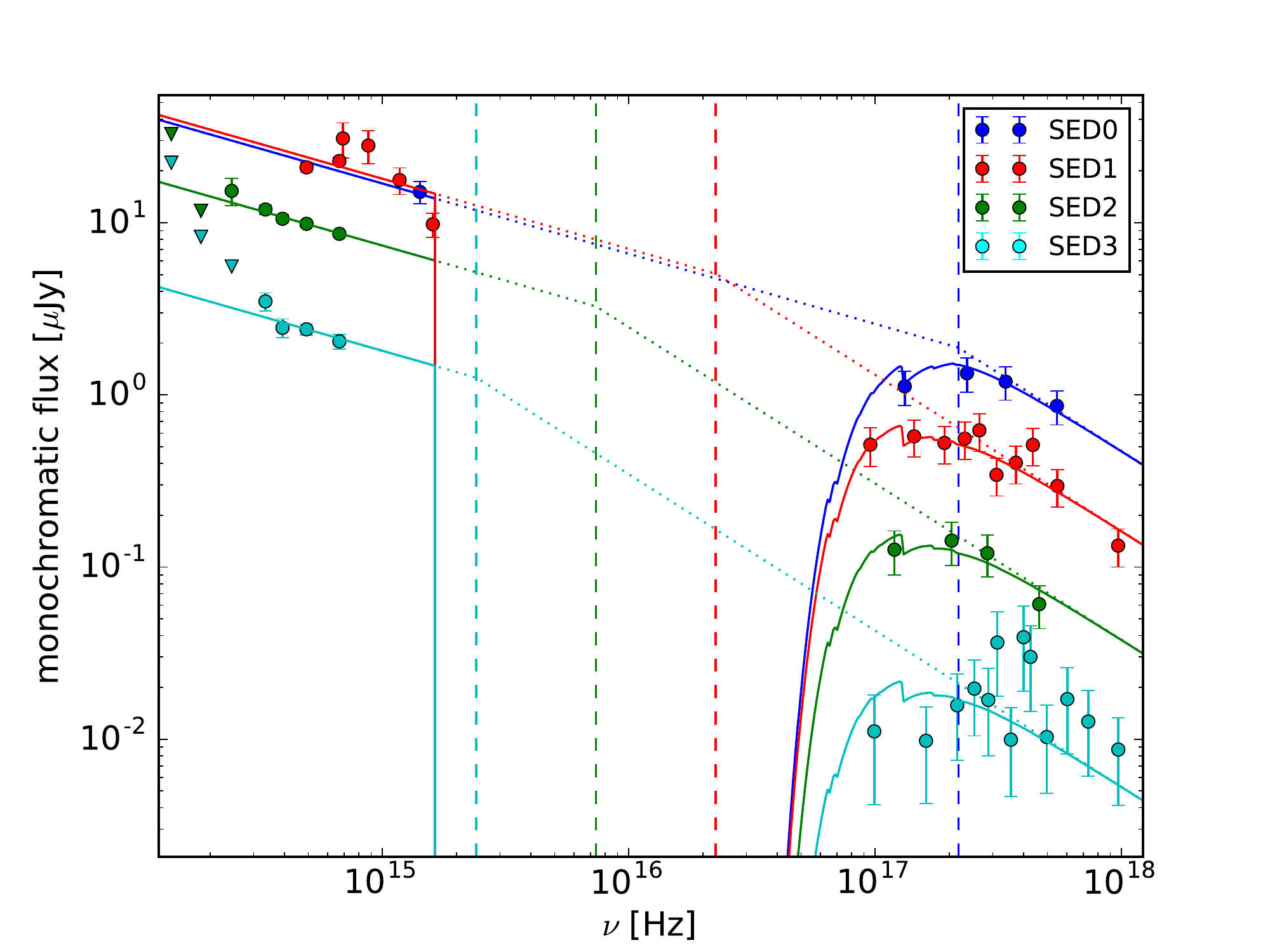}
   \caption{Broad band fit of 4 SEDs during the time bins indicated in Fig. \ref{Fig:LC150424A}.
   All 4 SEDs share the same spectral slopes. 
    {The solid lines are the folded model, 
   the dotted lines are the unfolded model.
   The vertical dashed lines mark the characteristic break frequency.}
   One can see a characteristic frequency evolving towards the low energy part of the spectrum.   
   }
   \label{Fig:150424A_SED1}
\end{figure}

The SED fits including the data and the unfolded model (just the power law, without extinction and absorption)
are shown in Fig. \ref{Fig:150424A_SED1}. 
All results of our analysis for $\alpha$ and $\beta$ are given in Tab.~\ref{Tab:PhenoAnalysis}.
A summary of the physical values from the spectral fitting is summarized in Tab.~\ref{Tab:AddData}.

\begin{figure}[ht]
  \centering
  \includegraphics[width=\hsize]{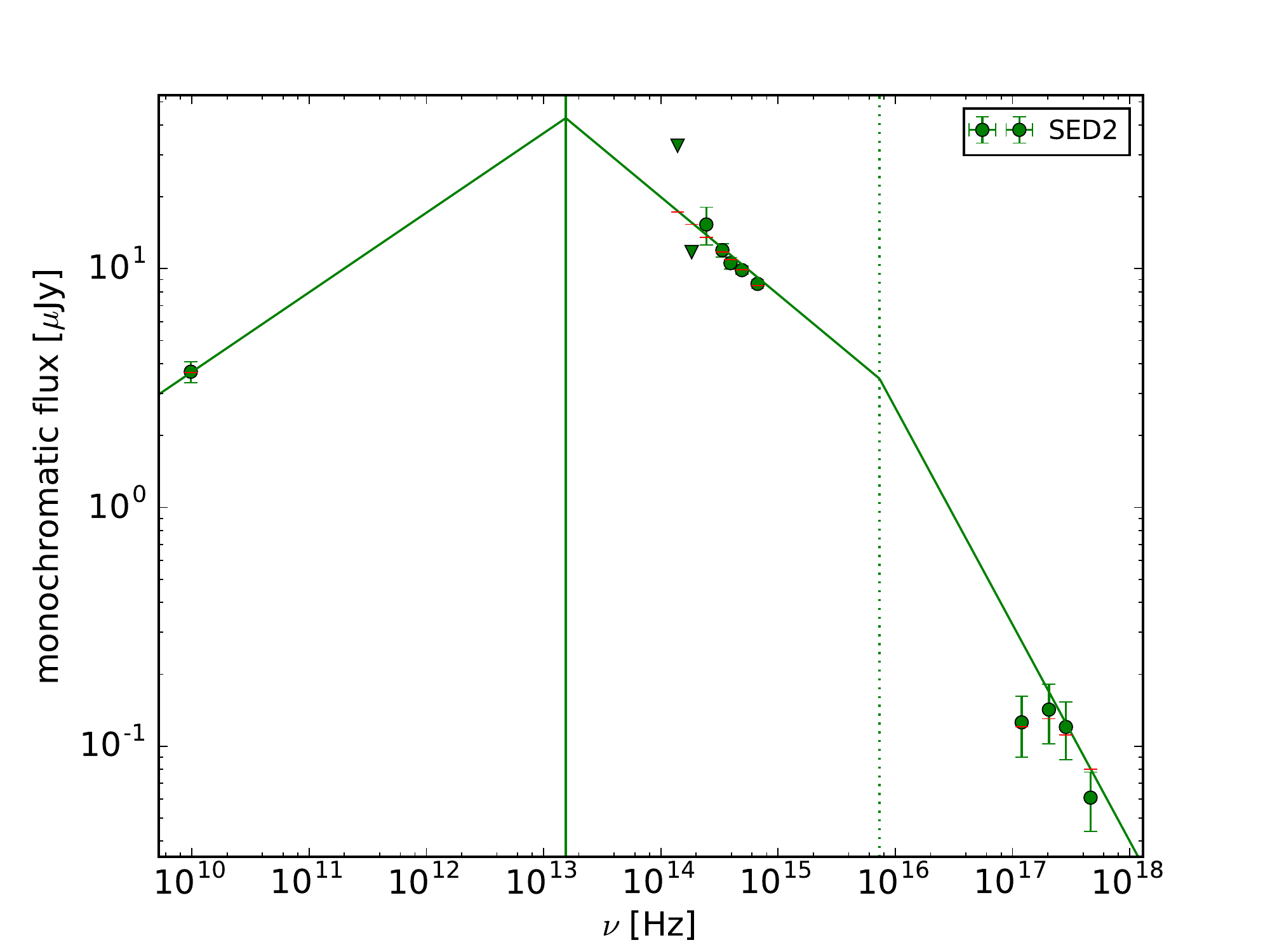}
   \caption{ {Unfolded} SED2 $t_{\mathrm{SED2}}=57900$ s after the trigger.
   The radio detection allows us to determine the peak frequency of the spectrum, 
   using a spectral slope $\beta=-1/3$ below the peak frequency,
   and the spectral slope derived in the multi SED fit. 
   }
   \label{Fig:150424A_SED2}
\end{figure}

The radio detection from \citet{Fong2015a} at time $t_{\mathrm{SED2}}=57900\,$s allows us to constrain the peak of the SED, and both characteristic frequencies 
(see Fig. \ref{Fig:150424A_SED2}).
For the spectral slope below the maximum frequency we assume the standard fireball $\beta=-1/3$, 
the other spectral slopes are the ones derived in the multi SED fit.

\begin{figure}[ht]
  \centering
  \includegraphics[width=\hsize]{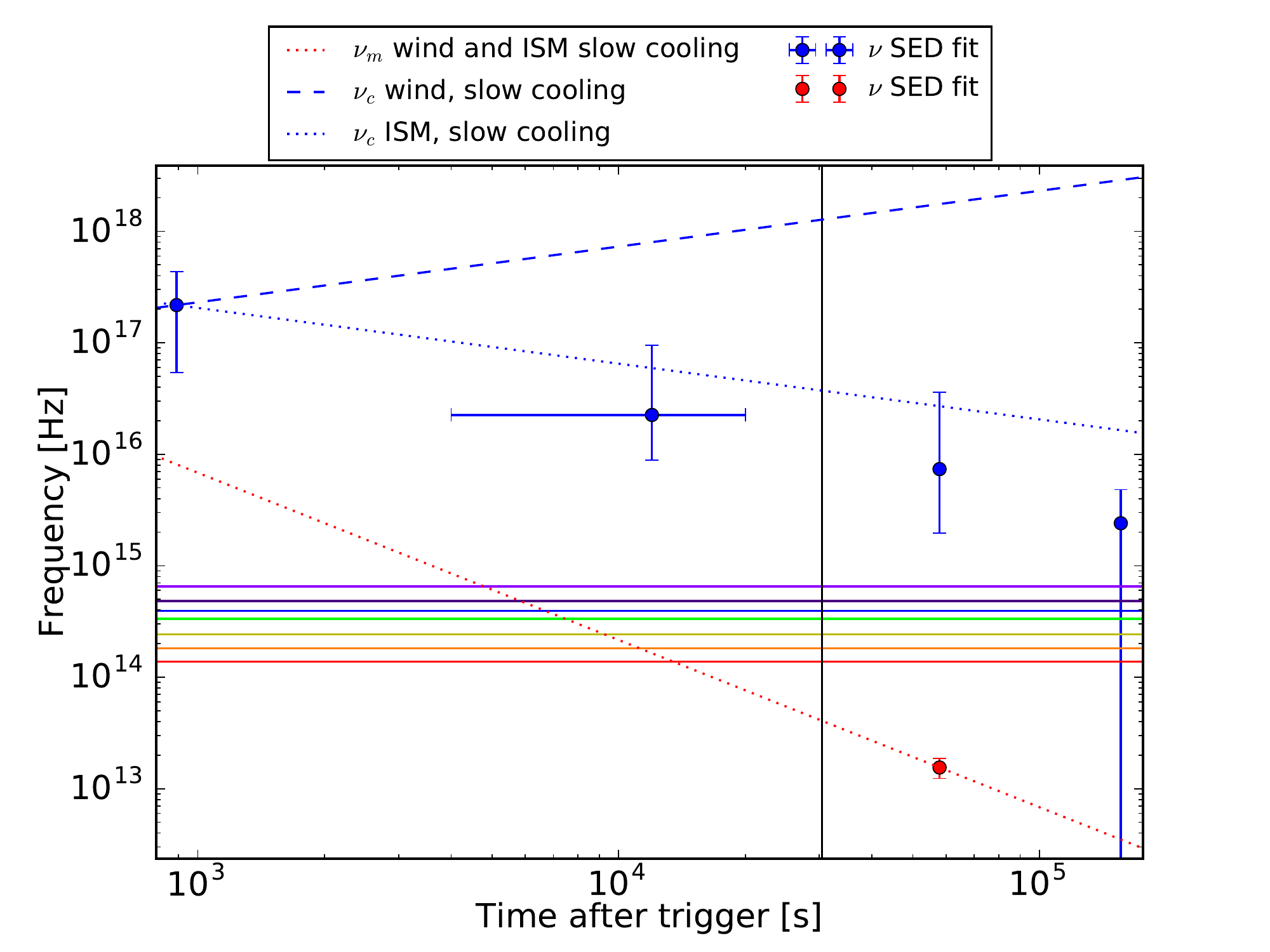}
   \caption{Evolution of the characteristic frequencies from the SED fit and the standard fireball model.
   The horizontal lines are the GROND bands.
   The vertical black line corresponds to the optical break time.  
   The blue data points are the break frequencies of the multi SED fit.
   The red data point is the peak frequency of the fit of SED2.
   The dashed and dotted lines are the evolutions of the characteristic frequencies according to different fireball scenarios. 
   }
   \label{Fig:nuEvolution}
\end{figure}

Fig. \ref{Fig:nuEvolution} shows the evolution of the characteristic frequencies based on our 4 SEDs,
 and the evolution of the characteristic frequencies according to some physical models,
 which we will discuss in Sec. \ref{sec:Discussion}.

\renewcommand{\arraystretch}{1.5}
\begin{table*}[ht]   
\caption{Summary of the phenomenological analysis.
         Temporal slopes $\alpha$ and 
         spectral slopes $\beta$ as defined in Eq. \ref{eq:ClosureRelations}.
         }  
\centering          
\begin{tabular}{c | c c | c c | c c | c}
      
  \multirow{2}{*}{GRB}    & \multirow{2}{*}{$t_{break, opt}$} & \multirow{2}{*}{$t_{break, xrt}$}   & \multicolumn{2}{c}{Optical}                  &\multicolumn{2}{c}{X-rays}  &  \\
                          &                                   &                                     &   $\alpha$         & $\beta$                 &     $\alpha$               &   $\beta$  &        \\
\hline
\hline
 \multirow{2}{*}{150424A} & \multirow{2}{*}{$30.4 \pm 0.9$ks} &  \multirow{2}{*}{$81.8 \pm 49.7$ks} & $0.00 \pm 0.01$   &  $0.41_{-0.07}^{+0.11}$  & $0.58 \pm 0.07$             & $0.91_{-0.07}^{+0.11}$      & $t<t_{break}$     \\
                          &                                   &                                     & $1.42 \pm 0.05$   &  $0.41_{-0.07}^{+0.11}$  & $1.66 \pm 0.26$             & $0.91_{-0.07}^{+0.11}$      & $t>t_{break}$     \\ 
\hline
                          &                                   &                                     & \multicolumn{2}{c}{smoothness: $15.0 \pm 13.8$}&\multicolumn{2}{|c|}{smoothness: $1.0 \pm 0.8$}            &      \\         
 \end{tabular}\label{Tab:PhenoAnalysis} 
\end{table*}
\renewcommand{\arraystretch}{1.0}

\begin{table*}   [ht]  
\caption{Additional data for the GRB }  
\centering          
\begin{tabular}{c |  c l}

                                 & 150424A                    & reference \\
\hline
Galactic E(B-V)[mag]             &  $0.051 \pm  0.002$      & \cite{Schlafly2011} \\
Galactic $N_H$ $[10^{21}cm^{-2}]$&  $0.60$                   & \cite{Kalberla2005}  \\
z                                &   $1.0 ^{+0.3}_{-0.2}$       & afterglow photometry          \\
$d_L [ 10^{28} cm ]$             &  2.1                       &  from z \\
Host E(B-V)  [mag]               & $0.0 ^{+1.5} _{-0.0}$      & SED fit                \\      
Host $N_H$  $[10^{22}cm^{-2}]$   & $0.04^{+0.24}_{-0.04}$     & SED fit          \\
$R_V$                            & 3.08 (MW)                  & assumption         \\
 \end{tabular}\label{Tab:AddData} 
\end{table*}

\subsection{Closure relations}\label{Sec:testCR}
Within the fireball model a GRB afterglow is explained as synchrotron radiation of shock-accelerated electrons
 from an ultra-relativistic outflow hitting the circumburst medium.
In a dynamical afterglow model one can link the temporal and spectral behavior of an afterglow over a set of so-called \textit{closure relations} between $\alpha$ and $\beta$.
They depend on the state and structure (isotropic or jet) of the outflow, the circumburst density profile, and the energy distribution of the electrons.

The dynamic of the relativistic outflow is influenced by the circumburst medium. 
 {For long GRBs}, two scenarios are usually considered.
First, the \textit{Inter-Stellar Medium (ISM) case}, where the circumburst medium is assumed to be homogeneous.
Second, the \textit{stellar wind case}, where the circumburst medium has radial density profile $n=n_0(R/R_0)^{-2}$, 
with a reference density $n_0$ and a reference radius $R_0$. 
 {For short GRBs the ISM case is expected.
 They are thought to be the result of a CBM,
  and a compact binary system does not produce a stellar wind during its lifetime.}

A synchrotron spectrum is conveniently characterized by characteristic frequencies:
The injection frequency $\nu_m$, which derives from the peak of the electron energy distribution,
and the cooling frequency $\nu_c$ above which the electrons lose a significant amount of energy via synchrotron cooling.
The \textit{slow-cooling} case is defined as $\nu_m<\nu_c$,
the \textit{fast-cooling} case is defined as $\nu_m>\nu_c$.
 
We compare the fitted $\alpha$ and $\beta$ to the theoretical closure relations collected by \citet{Racusin2009} 
(from \citealt{Zhang2004, Zhang2006, Dai2001, Panaitescu2006, Panaitescu2005}). 
We will abbreviate the closure relations with CR 1-14 from now (see Tab. 1 in \citet{Racusin2009} and Tab. \ref{Tab:ClosureRelations}).
We follow this scheme:
We use the $\beta$ value fitted to the data. 
Then we calculate $p(\beta)$. 
In \cite{Racusin2009} there are multiple ways to calculate $\alpha$.
They differ if $p>2$ or $p<2$, and if we want to consider energy injection.
We first calculate $\alpha(\beta)$ without energy injection.
If the calculated $\alpha$ is consistent with the $\alpha$ we fitted to the data, we consider it a "plausible scenario".
If not, we calculate the energy injection index $q(\alpha, \beta)$.
$q$ is defined using
\begin{equation}
L_{\mathrm{inj}}(t) = L_0 (t/t_b)^{-q}
\end{equation}
and is valid for $q\leqslant 1$.
$L_{\mathrm{inj}}$ is the luminosity injected into the blast wave.
The time $t$ and break time $t_b$ are given in the observer frame. 
$q=1$ is the impulsive injection case, and $q=0$ corresponds to a constant energy injection from e.g. a magnetar spin-down \citep{Zhang2006}.
We check all $\alpha-\beta$ pairs from the phenomenological analysis (Tab. \ref{Tab:PhenoAnalysis}).
In Tab. \ref{Tab:ClosureRelations} we list all closure relations that describe our afterglow,
 and the $q$ values if needed. 

In the SED fits we saw a break between the spectral regimes,
 so a set of closure relations can just be consistent with both spectral regimes
 if we find closure relations that lie on different sides of a characteristic frequency,
 and describe the same scenario.
After the break we just consider scenarios without energy injection to be plausible. 

We find only one scenario that describes both spectral regimes (optical and X-ray), before and after the break: 
A uniform non-spreading jet in an ISM environment. 
After the temporal break the scenario is consistent with the optical data (CR11, $\nu_m < \nu < \nu_c$) and the X-ray (CR12, $\nu > \nu_c$),
without the need for energy injection.
Before the temporal break, the optical data implies the need for energy injection with an injection index $q_{\mathrm{opt}}=0.06 \pm 0.07$, 
and the X-ray data imply a consistent $q_{x}= 0.14 \pm 0.11$. 

\subsection{Redshift and host}
\begin{figure}[ht]
  \centering
  \includegraphics[width=\hsize]{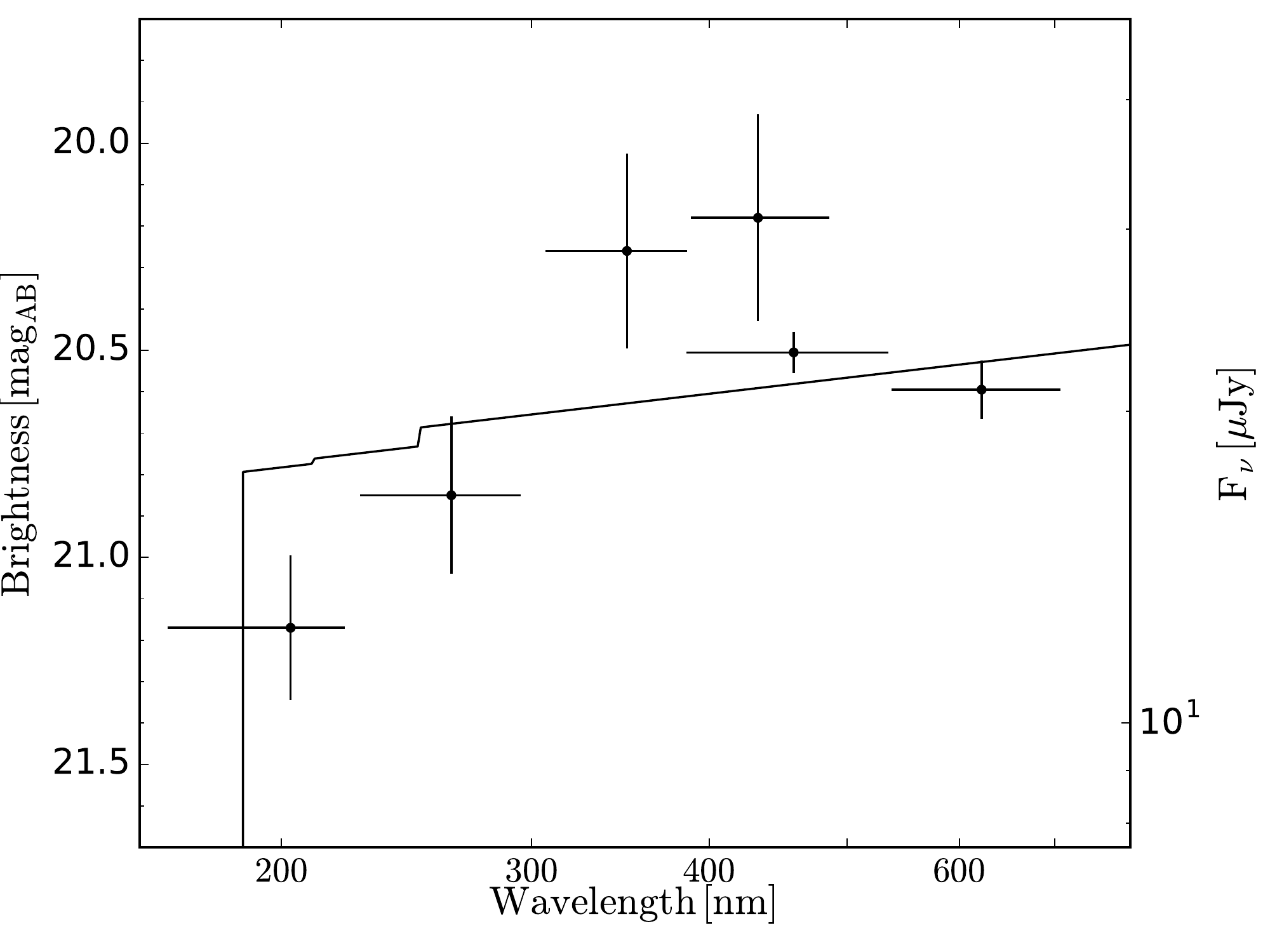}
   \caption{Best-fitting power-law template for the afterglow SED1. A  $z= 1.0 ^{+0.3}_{-0.2}$ is implied.
   }
   \label{Fig:PhotoZ}
\end{figure}

Using the extinction corrected SED1 (see Fig. \ref{Fig:PhotoZ}), 
we obtain the redshift from the  afterglow as described by \citet{Kruhler2011}, 
 and find $z= 1.0 ^{+0.3}_{-0.2}$.
Using a standard $\Lambda$CDM cosmology with 
$H_0=67.3\frac{km/s}{Mpc}$, $\Omega_m=0.315$, $\Omega_\Lambda=0.685$ 
\citep{PlanckCollaboration2014}, this corresponds to
a distance modulus of $44.2$ mag, and 
a luminosity distance of $D_L=2.1 \times 10^{28}$cm.
 
We note that  both, the size ($<$1\arcsec, corresponding to $<$8 kpc diameter)
and absolute luminosity (M$_B \sim -17.2\pm0.5$ mag) of the host are 
unusual for short-duration GRBs.

\subsection{The plausibility of a magnetar central engine}
\citet{Zhang2001} suggested a magnetar model as short GRB progenitor with a prolonged energy injection.
According to that model the magnetic field strength $B_{0p}$ at the poles of a magnetar is linked to its spin down luminosity $L_{\mathrm{sd}}$, 
and the initial spin period $P_{0}$ of the magnetar is linked to the spin-down time of the magnetar $\tau$:
\begin{equation}
B^2 _{0p, 15} = 4.20 \, I^2 _{45} R_6 ^{-6} L_{\mathrm{sd},49}   ^{-1} \tau ^{-2} _{\mathrm{sd}, 3}
\end{equation}
\begin{equation}
P^2 _{0,-3} = 2.05 \, I_{45}  L_{\mathrm{sd}, 49}  ^{-1} \tau^{-1} _{\mathrm{sd}, 3}
\end{equation}
with the moment of inertia $I_{45}$ in $10^{45} \mathrm{g\, cm}^2$. 
The spin down luminosity of the magnetar $L_{\mathrm{sd}, 49}$ is given in $10^{49} \mathrm{erg/s}$.
$B _{0p, 15}$ is in units of $10^{15}$ Gauss.
In the special case of an EE short GRB, $P_0$ corresponds to the spin period after EE \citep{Zhang2007, Gompertz2013}, rather than the spin down period when formed. 
$\tau_{\mathrm{sd}, 3}$ is the spin-down energy release time scale in units of $10^{3}\,$s, which corresponds to our plateau end time, respectively break time $t_{\mathrm{break,xrt}}$.
Following  \citet{Zhang2001} we adopt a radius of the neutron star $R_6 = 10^6$ cm and a neutron star mass  $m=1.4M_\odot$,
 which leads to $I=1.85 \times 10^{45} \mathrm{g \, cm}^2$.

The efficiency with which $L_{sd, 49}$ is converted to the afterglow luminosity in a specific band $X$ is 
\begin{equation}
L_X \equiv \eta_1 L_{\mathrm{BOL}} \equiv \eta_1  \eta_2  L_{\mathrm{inj}} \equiv \eta_{12}  L_{\mathrm{inj}}  \equiv \eta_{12}  L_{\mathrm{sd}}
\end{equation}
with the bolometric afterglow luminosity $L_{\mathrm{BOL}}$, the luminosity injected into the blast wave $L_{\mathrm{inj}}$, the observed luminosity $L_X$,
and the corresponding efficiency factors $\eta$.
We assume that all the magnetar spin-down luminosity is injected into the blast wave $L_{\mathrm{inj}}=L_{\mathrm{sd}}$.

Generally $\eta_{12}$  is time dependent, since the characteristic quantities of a synchrotron spectrum in the  fireball model also evolve with time.
For a constant energy injection $q=0$ and a $p \sim 2$ this time dependence is expected to be weak, 
in an ISM environment $k=0$ or wind environment $k=2$ (see fireball flux equation in e.g. \citealt{vanEerten2014}).

We use the break time $t_{\mathrm{break, xrt}}$ of the X-ray light curve, and SED2 to calculate the rest frame luminosity at that time. 
\cite{Gompertz2013, Rowlinson2014, Rea2015} approximate the bolometric luminosity with the 1-10000\,keV band, 
extrapolated from Swift data.
Using the \textit{lumin} command of \textit{Xspec} and a dummy response leads to a luminosity 
$L_{1-10000\,\mathrm{keV}} = 2.6 \times 10^{46} \mathrm{erg\, s}^{-1}$. 
When we integrate SED2 ($10^{-14}-10^{4}\,\mathrm{keV}$) we find a similar
$L_{0-10000\,\mathrm{keV}} = 2.7 \times 10^{46} \mathrm{erg\, s}^{-1}$.
Assuming a 10\% error in the luminosity, and $\eta_{12}=1$ this results in magnetar parameters
$B_{0p} = (0.9 \pm 0.5)\times 10^{15} G$ and $P_0 = 4.3 \pm 1.3$ms.

Since naturally $\eta_{12} < 1$ our values for $B_{0p}$ and $P_0$ therefore have to be seen as upper limits. 
Moreover, a $\eta_{12} = 1$ would mean that all the injected luminosity is radiated away immediately, 
and nothing goes into the kinetic energy of the outflow.
The fireball model therefore only can be self-consistent when $\eta_{12} << 1$.

Our $P_0$ lies above the mass-shedding limit $P_0=0.81$ms \citep{Lattimer2004}
below which a neutron star would be disrupted due to centrifugal forces.
For our fitted $t_{\mathrm{break, xrt}}$, the mass-shedding limit is reached when the $L_{\mathrm{sd,max}}>(7\pm4)\times 10^{47} \,\mathrm{erg \,s^{-1}}$.
This yields a  $\eta_{12} \gtrsim 4\% $.

The isotropic energy in the $\gamma$ band is 
\begin{equation}
E_{\gamma, \mathrm{iso}} =  f_\gamma  \times  D_L(z)^2 4 \pi \times (1+z)^{-1}
\end{equation}
with the $\gamma$-fluence $f_\gamma=15(1)\times 10^{-7} \,\mathrm{ erg\,cm^{-2}}$  
measured by the Swift/BAT \footnote{http://swift.gsfc.nasa.gov/archive/grb\_table.html} 
and the luminosity distance $D_L$, 
follows $E_{\gamma, \mathrm{iso}}=4.1 \times 10^{51} \mathrm{erg}$.
$E_{\gamma, \mathrm{iso}}$ is a proxy for the impulsive energy input into the the blast wave \citep{Granot2006}.
The prolonged energy injected is the luminosity at the end of the plateau times the length of the plateau.
 {Assuming an $\eta_{12}=1$}, the sum of both is the total energy of the blast wave
$E_{\mathrm{tot}} = E_{\gamma, \mathrm{iso}} + t_{\mathrm{break, xrt}}\times L_{0-10000\,\mathrm{keV}} =       (6.3 \pm 1.4) \times 10^{51}\,\mathrm{erg}$
 and does not exceed the limits for the maximal rotational energy of a proto magnetar of $1-2 \times 10^{53}\,$erg suggested by  \citet{Metzger2015a}.

\section{Discussion}\label{sec:Discussion}
\subsection{General description}

The X-ray light curve of GRB 150424A shows a steep decay from the prompt emission, followed by a smoothly broken power law.
GRB 150424A seems to be a typical EE short GRB \citep{Gompertz2013}.
This GRB becomes special since it is one of the rare cases with an early multi-epoch optical coverage,
 during which the  optical emission is basically constant for 8 hrs.
While the temporal break in the optical is very sharp, the break of the X-ray light curve is more smooth.  
Yet the breaks occur around the same time, which is a strong indicator that the underlying dynamics change at that point in time, 
i.e. the end of the optical plateau.

\begin{figure}[ht]
  \centering
  \includegraphics[width=1.1\hsize]{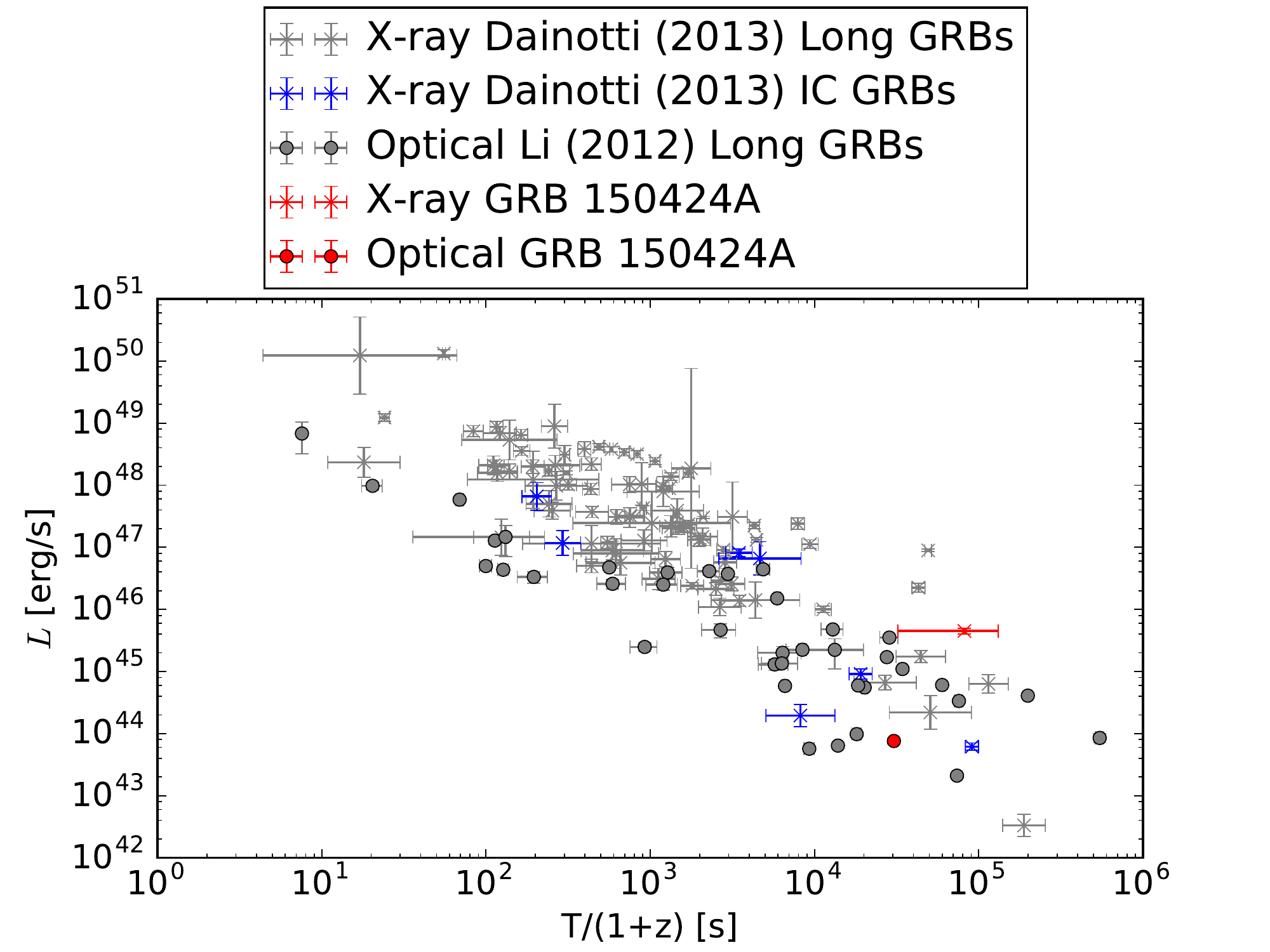}
   \caption{Plateau end time luminosity correlations.
   The X-ray data for long GRBs and EE short GRBs (tagged as Intermediate Class (IC) GRBs) is taken from \citet{Dainotti2013} (grey and blue crosses).
   The optical data comes from \citet{Li2012} (grey circles).
   The afterglow of GRB 150424A does not show any special behavior, whether in the optical (red circle), nor in the X-ray (red cross). 
   }
   \label{Fig:LtoLtx}
\end{figure}

Studies concerning the relation between the end time of a plateau and the luminosity at the end of the plateau \citep{Dainotti2008, Li2012, Dainotti2013}
show that the afterglow of GRB 150424A represents a "typical" shallow decay afterglow. 
As seen in Fig. \ref{Fig:LtoLtx} the afterglow does not have an outstanding position in the plateau end time - luminosity parameter space.
 {
The position of GRB 150424A in the parameter space is consistent with long GBRs, for the optical and X-ray data.
Comparing the position to other short GRBs, the afterglow has a slightly higher X-ray luminosity.
However, the short GRB sample lacks optical data, and is too small to conclude GRB 150424A to be an atypical short GRB.    
}

\subsection{Physical interpretation}
With an electron distribution index $\sim 2$, an optical spectral slope $\sim 1/2$ and an X-ray spectral slope $\sim 1$,
the spectral fits do not allow us to distinguish between a slow or fast cooling case, using the spectral shapes given by \citet{Sari1998}.

In Sec. \ref{Sec:testCR} we tested the derived spectral and temporal slopes with the most common closure relations
 (see Tab. \ref{Tab:ClosureRelations}).
 Here we present our findings.
In Sec. \ref{ISMslow} and Sec. \ref{windslow} we discuss two standard afterglow scenarios, 
and in Sec. \ref{windinjection} a scenario with energy injection involved.

\subsection{Scenario 1: ISM, slow cooling}\label{ISMslow}
The spectral shape of SED2 is compatible with slow cooling in an ISM, 
and would allow us to constrain $\nu_m$ (see Fig. \ref{Fig:150424A_SED2}). 
The temporal evolution of $\nu_c$ also does fit the observation (blue dotted line and blue resp. red dots in Fig. \ref{Fig:nuEvolution}).

The spectral shape from our SED fits shows that the X-ray data lies well above the cooling break.
However, the X-ray data by itself can only be explained by closure relations for slow cooling in an ISM
when the observed X-ray frequency would be $\nu_m < \nu_{\mathrm{X-ray}} < \nu_c$ (CR1),
with energy injection before the break and no energy injection after the break.
The SED fits and the closure relation CR1 for the X-ray light-curve therefore contradict each other.

\subsection{Scenario 2: slow cooling in wind without energy injection}\label{windslow}
\citet{Granot2002} give a temporal slope $\alpha=0$ for this scenario when the observer frequency $\nu < \nu_m$.
Therefore, the optical plateau could be explained if $\nu_m$ is above the optical frequencies. 
If we extrapolate from SED2 and follow the temporal evolution $\nu_m \propto t^{-3/2}$ we find that it would cross our observed bands at the time of SED1
(red dotted line in Fig. \ref{Fig:nuEvolution}), before the temporal break in the optical.
In our SED fits, 
SED0 has just one UVOT-white band data point, 
and for SED1 we had to use a wide time bin for the optical part of the SED.
A $\nu_m$ crossing at that point in time may therefore not be detectable by our spectral fitting,
but it should coincide with the end of the plateau.

Before the break the X-ray data is not consistent with this scenario (to fulfill CR5 or CR6 in Tab. \ref{Tab:ClosureRelations} energy injection has to be accounted for).
After the break it is consistent when $\nu_m < \nu_{\mathrm{X-ray}} < \nu_c$ (CR5).
However, our SED fit shows that the fitted $\nu_c$ is well below the X-ray (blue dots in Fig. \ref{Fig:nuEvolution}), 
and that the evolution of the break frequency does not follow the predicted evolution for $\nu_c$ in this scenario
 (see blue dashed line and blue data points in Fig. \ref{Fig:nuEvolution}).

\subsection{Scenario 3: uniform nonspreading jet in ISM with energy injection}\label{windinjection}
The closure relations for a uniform nonspreading jet in an ISM medium with slow cooling
are valid for all temporal and spectral regimes (the optical is consistent with CR11, the X-ray is consistent with CR12).
Before the temporal break energy injection is needed, after the temporal break it is consistent with both spectral regimes without energy injection.
We derived an efficiency to convert the spin-down luminosity of the magnetar to the total afterglow luminosity $\eta_{12} \gtrsim 4 \%$.
Models that assume an adiabatic blast wave, like the one in we used in this scenario 3, harmonize with a $4\%$ loss due to radiation. 

We derived the energy injection index $q$ independently from the optical and the X-ray data.
Both $q$ values are consistent with a constant energy injection (q=0), as expected from a magnetar.

In this scenario 3, the jet nature of the outflow is apparent right from the beginning of the light curve,
 which means that the jet break already has occurred before the observations began, 
 before the first optical data point $ \lesssim 200\,$s. 
 This is more than one order of magnitude earlier than the earliest short GRB jet break measurement (GRB 090426A, $\sim 35000\,$s, \citealt{NicuesaGuelbenzu2011}, although it is debated if GRB 090426A is actually a short GRB), 
 or long GRB measurement (GRB 120729A $\sim 9500\,$s, \citealt{Cano2014a}),
 and would imply a very narrow jet opening angle.
 The narrower the jet opening angle, the lower the beaming corrected total energy output, given an isotropic energy equivalent.

\section{Conclusions}\label{sec:Conclusions}
We presented multi-band data with a uniquely high temporal and spectral coverage of the EE short GRB 150424A.
We did a phenomenological analysis and interpreted them in the context of the fireball model.
We found 3 scenarios that explain parts of the temporal and spectral behavior of the afterglow.
1. Slow cooling in an ISM 
2. Slow cooling in wind without energy injection
3. A uniform nonspreading jet in an ISM medium with energy injection.
We found that typical standard scenarios of GRB afterglows,
i.e. slow cooling in ISM or wind environment, are not able to explain our data. 

In contrast, a uniform, non-spreading jet expanding into an ISM medium 
and re-powered for $\approx 10^4\,$s with additional constant energy injection can explain the data self consistently. 
For a magnetar as supplier of this prolonged energy injection
an efficiency $\eta_{12} \gtrsim 4\% $ in converting the spin-down luminosity of the magnetar $L_{\mathrm{sd}}$ to the afterglow luminosity has to be assumed. 

The unique and very-long-duration energy injection provides, 
within a factor of 2, 
a similar energy input into the surroundings as the prompt GRB emission.
That the jet behaviour is apparent from very early times on, however, 
implies an extremely narrow jet opening angle.
 The narrower the jet opening angle, 
the smaller the beaming corrected total energy output.
Yet, even the total isotropic equivalent energy release is a factor of 20 below the maximum possible for a magnetar.

\begin{acknowledgements}
HJvE was supported by the Alexander von Humboldt foundation when this work was done. 

DAK acknowledges financial support from MPE, from TLS, from the 
Spanish research project AYA 2014-58381-P, and from Juan de la Cierva 
Incorporaci\'on fellowships IJCI-2015-26153 and IJCI-2014-21669. 

PS, TWC, JFG, MT acknowledge support through the Sofja Kovalevskaja Award from the Alexander von Humboldt Foundation of Germany.

SK and ANG acknowledge support by DFG grant Kl 766/16-1.

SS acknowledges support by the Th\"uringer Ministerium f\"ur Bildung, Wissenschaft und Kultur under FKZ 12010-514.
 
Part of the funding for GROND (both hardware as well as personnel) was generously granted from the Leibniz-Prize to Prof. G. Hasinger (DFG grant HA 1850/28-1).

This work made use of data supplied by the UK {\it Swift} Science Data Centre at the University of Leicester. 

Based on observations made with the NASA/ESA Hubble Space Telescope, 
obtained from the Data Archive at the Space Telescope Science Institute, 
which is operated by the Association of Universities for Research in Astronomy, Inc., 
under NASA contract NAS 5-26555. 
These observations are associated with program $\#$ 13830 Pi'ed by N. Tanvir.

\end{acknowledgements}


\bibliographystyle{aa}
\newcommand\aip{AIP }
\newcommand\asp{ASP }
\bibliography{Paper2}

\begin{appendix}

\section{Additional tables}

\renewcommand{\arraystretch}{1.5}
\begin{table*}[p]   
\caption{All closure relations that are consistent with the data. The energy injection index $q$ is given where needed. }    
\begin{tabular}{l | c c | r r  }
CR &  scenario                                      &  spectral regime          &  $p$                        & q   \\
\hline
\multicolumn{5}{c}{  X-ray before break } \\
\hline
1  & ISM, slow cooling                              & $\nu_m < \nu < \nu_c    $ & $2.82^{+0.21 }_{-0.13 }$ & $0.46 \pm 0.10$ \\
2  & ISM, slow cooling                              & $\nu > \nu_c            $ & $1.82^{+0.21 }_{-0.13 }$ & $0.71 \pm 0.17$ \\
4  & ISM, fast cooling                              & $\nu > \nu_m            $ & $1.82^{+0.21 }_{-0.13 }$ & $0.71 \pm 0.17$ \\
5  & wind, slow cooling                             & $\nu_m < \nu < \nu_c    $ & $2.82^{+0.21 }_{-0.13 }$ & $-0.34 \pm 0.12$ \\
6  & wind, slow cooling                             & $\nu > \nu_c            $ & $1.82^{+0.21 }_{-0.13 }$ & $-1.39 \pm 0.08$ \\
8  & wind, fast cooling                             & $\nu > \nu_m            $ & $1.82^{+0.21 }_{-0.13 }$ & $0.71 \pm 0.17$ \\
9  & uniform jet (spreading), slow cooling          & $\nu_m < \nu < \nu_c    $ & $2.82^{+0.21 }_{-0.13 }$ & $0.32 \pm 0.05$ \\
10 & uniform jet (spreading), slow cooling          & $\nu > \nu_c            $ & $1.82^{+0.21 }_{-0.13 }$ & $-2.08 \pm 0.32$ \\
11 & ISM, uniform jet (nonspreading), slow cooling  & $\nu_m < \nu < \nu_c    $ & $2.82^{+0.21 }_{-0.13 }$ & $0.10 \pm 0.08$ \\
12 & ISM, Uniform jet (nonspreading), slow cooling  & $\nu > \nu_c            $ & $1.82^{+0.21 }_{-0.13 }$ & $0.14 \pm 0.11$ \\

13 & wind, uniform jet (nonspreading), slow cooling & $\nu_m < \nu < \nu_c    $ & $2.82^{+0.21 }_{-0.13 }$ & $-0.22 \pm 0.08$ \\
14 & wind, uniform jet (nonspreading), slow cooling & $\nu > \nu_c            $ & $1.82^{+0.21 }_{-0.13 }$ & $0.46 \pm 0.10$ \\
\hline
\multicolumn{5}{c}{ X-ray after break } \\
\hline
1  & ISM, slow cooling                              & $\nu_m < \nu < \nu_c    $ & $2.82^{+0.21 }_{-0.13 }$ & no energy injection needed \\
5  & wind, slow cooling                             & $\nu_m < \nu < \nu_c    $ & $2.82^{+0.21 }_{-0.13 }$ & no energy injection needed \\
6  & wind, slow cooling                             & $\nu > \nu_c            $ & $1.82^{+0.21 }_{-0.13 }$ & $-0.26 \pm 0.29$ \\
9  & uniform jet (spreading), slow cooling          & $\nu_m < \nu < \nu_c    $ & $2.82^{+0.21 }_{-0.13 }$ & $0.87 \pm 0.14$ \\
11 & ISM, uniform jet (nonspreading), slow cooling  & $\nu_m < \nu < \nu_c    $ & $2.82^{+0.21 }_{-0.13 }$ & $0.74 \pm 0.17$ \\
12 & ISM, uniform jet (nonspreading), slow cooling  & $\nu > \nu_c            $ & $1.82^{+0.21 }_{-0.13 }$ & no energy injection needed \\
13 & wind, uniform jet (nonspreading), slow cooling & $\nu_m < \nu < \nu_c    $ & $2.82^{+0.21 }_{-0.13 }$ & $0.52 \pm 0.20$ \\
14 & wind, uniform jet (nonspreading), slow cooling & $\nu > \nu_c            $ & $1.82^{+0.21 }_{-0.13 }$ & no energy injection needed \\
\hline
\hline
\multicolumn{5}{c}{ optical before break } \\
\hline
1  & ISM, slow cooling                              & $\nu_m < \nu < \nu_c    $ & $1.82^{+0.21 }_{-0.13 }$ & $0.49 \pm 0.11$ \\
2  & ISM, slow cooling                              & $\nu > \nu_c            $ & $0.82^{+0.21 }_{-0.13 }$ & $0.84 \pm 0.21$ \\
3  & ISM, fast cooling                              & $\nu_c < \nu < \nu_m    $ & $na   ^{+na   }_{-na   }$ & $0.74 \pm 0.08$ \\
4  & ISM, fast cooling                              & $\nu > \nu_m            $ & $0.82^{+0.21 }_{-0.13 }$ & $0.84 \pm 0.21$ \\
5  & wind, slow cooling                             & $\nu_m < \nu < \nu_c    $ & $1.82^{+0.21 }_{-0.13 }$ & $-0.58 \pm 0.11$ \\
6  & wind, slow cooling                             & $\nu > \nu_c            $ & $0.82^{+0.21 }_{-0.13 }$ & $-2.00 \pm 0.01$ \\
7  & wind, fast cooling                             & $\nu_c < \nu < \nu_m    $ & $na   ^{+na   }_{-na   }$ & $0.58 \pm 0.11$ \\
8  & wind, fast cooling                             & $\nu > \nu_m            $ & $0.82^{+0.21 }_{-0.13 }$ & $0.84 \pm 0.21$ \\
9  & uniform jet (spreading), slow cooling          & $\nu_m < \nu < \nu_c    $ & $1.82^{+0.21 }_{-0.13 }$ & $0.12 \pm 0.03$ \\
10 & uniform jet (spreading), slow cooling          & $\nu > \nu_c            $ & $0.82^{+0.21 }_{-0.13 }$ & $-4.00 \pm 0.04$ \\
11 & ISM, uniform jet (nonspreading), slow cooling  & $\nu_m < \nu < \nu_c    $ & $1.82^{+0.21 }_{-0.13 }$ & $0.06 \pm 0.07$ \\
12 & ISM, Uniform jet (nonspreading), slow cooling  & $\nu > \nu_c            $ & $0.82^{+0.21 }_{-0.13 }$ & $0.10 \pm 0.12$ \\
13 & wind, uniform jet (nonspreading), slow cooling & $\nu_m < \nu < \nu_c    $ & $1.82^{+0.21 }_{-0.13 }$ & $-0.34 \pm 0.07$ \\
14 & wind, uniform jet (nonspreading), slow cooling & $\nu > \nu_c            $ & $0.82^{+0.21 }_{-0.13 }$ & $0.49 \pm 0.11$ \\
\hline
\multicolumn{5}{c}{ optical after break } \\
\hline
6  & wind, slow cooling                             & $\nu > \nu_c            $ & $0.82^{+0.21 }_{-0.13 }$ & $0.02 \pm 0.16$ \\
11 & ISM, uniform jet (nonspreading), slow cooling  & $\nu_m < \nu < \nu_c    $ & $1.82^{+0.21 }_{-0.13 }$ & no energy injection needed \\
13 & wind, uniform jet (nonspreading), slow cooling & $\nu_m < \nu < \nu_c    $ & $1.82^{+0.21 }_{-0.13 }$ & $0.84 \pm 0.13$ \\
\end{tabular}\label{Tab:ClosureRelations}
\end{table*}
\renewcommand{\arraystretch}{1.0}

\onecolumn

\begin{table*}[p]   
\caption{AB magnitudes of comparison stars. 
 They do not include the systematical errors of the calibration:
  $g'=0.03$ mag, $r'=0.03$ mag,  $i'=0.04$ mag, $z'=0.04$ mag, $J=0.05$ mag, $H=0.05$ mag, $K=0.07$ mag
    }  
\centering          
\begin{tabular}{l | r r | c c c c c c c}
 \# & RA      & Dec.         & $g'$ & $r'$ & $i'$ & $z'$ & $J$ & $H$ & $K_S$ \\
\hline
I   & 152.315 & -26.628 & $20.88 \pm 0.02$ & $19.96 \pm 0.01$ & $19.46 \pm 0.01$ & $19.19 \pm 0.01$ & $17.72 \pm 0.02$ & $17.05 \pm 0.03$ & $16.50 \pm 0.07$ \\ 
II  & 152.297 & -26.628 & $22.40 \pm 0.06$ & $21.58 \pm 0.03$ & $21.26 \pm 0.04$ & $21.10 \pm 0.05$ & $19.82 \pm 0.15$ & $18.98 \pm 0.11$ &  na          \\
III & 152.299 & -26.635 &    na             & $21.10 \pm 0.01$ & $19.91 \pm 0.01$ & $19.39 \pm 0.01$ & $18.08 \pm 0.03$ & $17.43 \pm 0.03$ & $17.25 \pm 0.14$ \\

 \end{tabular}\label{Tab:ComparisonStars} 
\end{table*}

\longtab{
\begin{longtable}{r r r r r r}
\caption{Photometry of GRB 150424A. Upper limits are flagged with "UL"}\\
\hline
\hline
Time  & Time error & mag & mag error & band & instrument \\
\hline
\label{Tab:photometry}
\endfirsthead
\caption{Continued.} \\
\hline
Time & Time error & AB mag & mag error & band & instrument \\
\hline
\endhead
\hline
\endfoot
\hline
\endlastfoot
57903 & 2312 & 21.76 & 0.04 & $g'$ & GROND \\
62645 & 2325 & 21.90 & 0.03 & $g'$ & GROND \\
67277 & 2220 & 21.99 & 0.04 & $g'$ & GROND \\
156123 & 4528 & 23.32 & 0.10 & $g'$ & GROND \\
238900 & 5428 & 23.6 & UL    & $g'$ & GROND \\
323218 & 2930 & 24.5 & UL & $g'$ & GROND \\
411767 & 5450 & 24.9 & UL & $g'$ & GROND \\
582978 & 3189 & 23.4 & UL & $g'$ & GROND \\
670665 & 5245 & 23.9 & UL & $g'$ & GROND \\
842737 & 4961 & 24.2 & UL & $g'$ & GROND \\
57903 & 2312 & 21.55 & 0.03 & $r'$ & GROND \\
62645 & 2325 & 21.60 & 0.03 & $r'$ & GROND \\
67277 & 2220 & 21.80 & 0.03 & $r'$ & GROND \\
156582 & 4986 & 23.08 & 0.07 & $r'$ & GROND \\
238900 & 5428 & 24.4 & UL & $r'$ & GROND \\
323218 & 2930 & 23.92 & 0.22 & $r'$ & GROND \\
411767 & 5450 & 25.0 & UL & $r'$ & GROND \\
582978 & 3189 & 23.9 & UL & $r'$ & GROND \\
670891 & 5471 & 24.1 & UL & $r'$ & GROND \\
842962 & 5186 & 24.5 & UL & $r'$ & GROND \\
57903 & 2312 & 21.45 & 0.05 & $i'$ & GROND \\
62645 & 2325 & 21.44 & 0.04 & $i'$ & GROND \\
67374 & 2316 & 21.58 & 0.05 & $i'$ & GROND \\
156582 & 4986 & 23.03 & 0.13 & $i'$ & GROND \\
238900 & 5428 & 23.9 & UL & $i'$ & GROND \\
323218 & 2930 & 24.1 & UL & $i'$ & GROND \\
411767 & 5450 & 24.6 & UL & $i'$ & GROND \\
582978 & 3189 & 23.6 & UL & $i'$ & GROND \\
670891 & 5471 & 23.8 & UL & $i'$ & GROND \\
842962 & 5186 & 24.3 & UL & $i'$ & GROND \\
57903 & 2312 & 21.29 & 0.06 & $z'$ & GROND \\
62645 & 2325 & 21.40 & 0.05 & $z'$ & GROND \\
67374 & 2316 & 21.52 & 0.06 & $z'$ & GROND \\
156355 & 4760 & 22.62 & 0.12 & $z'$ & GROND \\
238900 & 5428 & 23.6 & UL & $z'$ & GROND \\
323218 & 2930 & 23.8 & UL & $z'$ & GROND \\
411767 & 5450 & 24.4 & UL & $z'$ & GROND \\
582978 & 3189 & 23.3 & UL & $z'$ & GROND \\
670891 & 5471 & 23.0 & UL & $z'$ & GROND \\
842962 & 5186 & 24.0 & UL & $z'$ & GROND \\
57929 & 2338 & 20.98 & 0.19 & $J$ & GROND \\
62670 & 2351 & 21.22 & 0.19 & $J$ & GROND \\
67399 & 2343 & 21.38 & 0.23 & $J$ & GROND \\
156606 & 5011 & 22.1 & UL & $J$ & GROND \\
238924 & 5452 & 22.0 & UL & $J$ & GROND \\
323331 & 3043 & 21.9 & UL & $J$ & GROND \\
411791 & 5475 & 22.4 & UL & $J$ & GROND \\
670915 & 5494 & 22.1 & UL & $J$ & GROND \\
842987 & 5211 & 22.4 & UL & $J$ & GROND \\
156606 & 5011 & 21.6 & UL & $H$ & GROND \\
238924 & 5452 & 21.6 & UL & $H$ & GROND \\
323331 & 3043 & 21.5 & UL & $H$ & GROND \\
411791 & 5475 & 22.0 & UL & $H$ & GROND \\
670915 & 5494 & 21.6 & UL & $H$ & GROND \\
842987 & 5211 & 21.7 & UL & $H$ & GROND \\
156834 & 5239 & 20.5 & UL & $K_S$ & GROND \\
238924 & 5452 & 20.0 & UL & $K_S$ & GROND \\
323242 & 2954 & 19.9 & UL & $K_S$ & GROND \\
411791 & 5475 & 20.4 & UL & $K_S$ & GROND \\
670915 & 5494 & 20.5 & UL & $K_S$ & GROND \\
843441 & 5665 & 20.9 & UL & $K_S$ & GROND \\
436 & 123 & 21.3 & UL & $u$ & UVOT \\
2887 & 108 & 20.8 & UL & $u$ & UVOT \\
6396 & 98 & 20.56 & 0.23 & $u$ & UVOT \\
3404 & 329 & 20.91 & 0.17 & $u$ & UVOT \\
45591 & 433 & 21.3 & UL & $u$ & UVOT \\
99360 & 305 & 21.0 & UL & $u$ & UVOT \\
185608 & 300 & 20.8 & UL & $u$ & UVOT \\
795276 & 284 & 21.3 & UL & $u$ & UVOT \\
1394595 & 4354 & 22.1 & UL & $u$ & UVOT \\
664 & 19 & 18.9 & UL & $b$ & UVOT \\
5884 & 197 & 20.42 & 0.25 & $b$ & UVOT \\
3634 & 216 & 20.9 & UL & $b$ & UVOT \\
16166 & 443 & 20.5 & UL & $b$ & UVOT \\
16166 & 443 & 21.0 & UL & $b$ & UVOT \\
129408 & 590 & 20.7 & UL & $b$ & UVOT \\
82 & 5 & 17.5 & UL & $v$ & UVOT \\
738 & 19 & 18.0 & UL & $v$ & UVOT \\
4347 & 98 & 19.1 & UL & $v$ & UVOT \\
5782 & 98 & 19.6 & UL & $v$ & UVOT \\
3261 & 216 & 19.8 & UL & $v$ & UVOT \\
12078 & 295 & 19.9 & UL & $v$ & UVOT \\
20295 & 589 & 20.1 & UL & $v$ & UVOT \\
34741 & 442 & 19.9 & UL & $v$ & UVOT \\
160892 & 588 & 19.9 & UL & $v$ & UVOT \\
2773 & 108 & 20.9 & UL & $uvw1$ & UVOT \\
17934 & 329 & 21.27 & 0.19 & $uvw1$ & UVOT \\
42244 & 560 & 22.2 & UL & $uvw1$ & UVOT \\
92657 & 358 & 22.1 & UL & $uvw1$ & UVOT \\
177963 & 461 & 21.8 & UL & $uvw1$ & UVOT \\
746677 & 1086 & 22.4 & UL & $uvw1$ & UVOT \\
1310529 & 5447 & 22.6 & UL & $uvw1$ & UVOT \\
714 & 19 & 19.7 & UL & $uvw2$ & UVOT \\
3349 & 106 & 21.2 & UL & $uvw2$ & UVOT \\
3146 & 126 & 21.9 & UL & $uvw2$ & UVOT \\
11320 & 443 & 21.65 & 0.17 & $uvw2$ & UVOT \\
29251 & 1618 & 21.89 & 0.10 & $uvw2$ & UVOT \\
129298 & 554 & 22.6 & UL & $uvw2$ & UVOT \\
215616 & 640 & 22.5 & UL & $uvw2$ & UVOT \\
431143 & 3532 & 23.1 & UL & $uvw2$ & UVOT \\
1295417 & 6031 & 23.2 & UL & $uvw2$ & UVOT \\
762 & 19 & 19.1 & UL & $uvm2$ & UVOT \\
5269 & 197 & 21.2 & UL & $uvm2$ & UVOT \\
3376 & 206 & 21.7 & UL & $uvm2$ & UVOT \\
72406 & 470 & 21.9 & UL & $uvm2$ & UVOT \\
173186 & 1105 & 22.3 & UL & $uvm2$ & UVOT \\
516450 & 2747 & 22.5 & UL & $uvm2$ & UVOT \\
1224527 & 5590 & 22.4 & UL & $uvm2$ & UVOT \\
174 & 74 & 21.30 & 0.23 & $white$ & UVOT \\
601 & 10 & 20.7 & UL & $white$ & UVOT \\
890 & 83 & 20.94 & 0.16 & $white$ & UVOT \\
5371 & 98 & 20.96 & 0.15 & $white$ & UVOT \\
8481 & 418 & 20.93 & 0.08 & $white$ & UVOT \\
10563 & 295 & 20.87 & 0.09 & $white$ & UVOT \\
16926 & 295 & 21.09 & 0.10 & $white$ & UVOT \\
32126 & 1053 & 21.02 & 0.06 & $white$ & UVOT \\
292642 & 1707 & 22.8 & UL & $white$ & UVOT \\
303903 & 1184 & 22.7 & UL & $white$ & UVOT \\
314852 & 534 & 22.2 & UL & $white$ & UVOT \\
335542 & 427 & 22.3 & UL & $white$ & UVOT \\
358922 & 1178 & 22.6 & UL & $white$ & UVOT \\
576743 & 2047 & 24.96 & 0.25 & $F125W$ & HST \\
800368 & 2197 & 25.53 & 0.26 & $F125W$ & HST \\
1203117 & 811 & 26.01 & 0.26 & $F125W$ & HST \\
579662 & 811 & 24.73 & 0.25 & $F160W$ & HST \\
580609 & 75 & 24.55 & 0.25 & $F160W$ & HST \\
803437 & 811 & 25.26 & 0.25 & $F160W$ & HST \\
1206264 & 2274 & 25.57 & 0.26 & $F160W$ & HST \\
573431 & 1092 & 26.03 & 0.26 & $F606W$ & HST \\
796906 & 1092 & 26.25 & 0.26 & $F606W$ & HST \\
5604 & 180 & 20.77 & 0.03 & $g^\prime$ & LRIS \\
5616 & 180 & 20.66 & 0.04 & $R_c$ & LRIS \\
\end{longtable}
}

\end{appendix}
\end{document}